\documentstyle[floats,tighten,preprint,aps,epsfig]{revtex}

\begin{document}

\title{ Parity-violating longitudinal response
\thanks{This work is supported in part by
funds provided by the U.S. Department of Energy (D.O.E.) under cooperative
agreement \#DE-FC02-94ER40818    .}
}
\author{
 {M.B. Barbaro}$^a$, {A. De Pace}$^a$, {T.W. Donnelly}$^b$
 and A. Molinari$^a$
}
\address{
 $^a$ Dipartimento di Fisica Teorica dell'Universit\`a
 di Torino and \\
 Istituto Nazionale di Fisica Nucleare, Sezione di Torino, \\
 via P.Giuria 1, I-10125 Torino, Italy \\
 $^b$ Center for Theoretical Physics, \\
 Laboratory for Nuclear Science and Department of Physics, \\
 Massachusetts Institute of Technology, Cambridge, MA 02139, USA
}

\maketitle

\begin{abstract}
The longitudinal quasielastic parity-violating electron scattering response
is explored within the context of a model that builds antisymmetrized
RPA-HF correlations on a relativistic Fermi gas basis. The large
sensitivity to nuclear dynamics of this observable, found in previous studies
where only pionic
correlations were included, is shown to survive in the present model where
the effects from $\pi$, $\rho$, $\sigma$ and $\omega$ exchange in a version
of the Bonn potential are incorporated. Through an intricate
diagrammatic cancellation/filtration mechanism the longitudinal
parity-violating response turns out to be close to the one obtained in
first-order perturbation theory with the pion alone. Finally, in accord
with our previous work, the parity-violating response is seen to display
appreciable sensitivity to the electric strangeness content of the nucleon,
especially at high momentum transfer.
\end{abstract}

\vfill
\begin{center}
Submitted to: {\em Nuclear Physics A}
\end{center}

\vfill

CTP\#2424 \hfill August 1995

\eject

\section{Introduction}
\label{sec:intro}

The unusual nature of the quasielastic parity-violating (pv)
longitudinal response $R^L_{AV}$ was first explored within the context of
the pionic-correlated Fermi gas model in our previous work
\cite{Don92,Alb93,Bar94}. The enhanced isospin-correlation sensitivity
found in that work can be traced back to the critical balance between
the isoscalar ($\tau=0$) and isovector ($\tau=1$) components in the
longitudinal response functions. Indeed, on the one hand, in the
electromagnetic (em) charge response,
\begin{equation}
  R^L(q,\omega) = R^L(\tau=0) + R^L(\tau=1),
\label{eq:RL}
\end{equation}
the isoscalar and isovector pieces enter with the same sign and with the
same norm in the independent-particle model, if the effects stemming from
the nucleon form factors are neglected. In the weak-neutral current case,
on the other hand, one has
\begin{equation}
  R^L_{AV}(q,\omega) = a_A \left[\beta_V^{(0)} R^L(\tau=0) + \beta_V^{(1)}
    R^L(\tau=1)\right],
\label{eq:RLAV}
\end{equation}
where $a_A=-1$ and the other constants appearing in the above relation
are given by \cite{Don92}
\begin{mathletters}
\begin{eqnarray}
  \beta_V^{(0)} &=& 1 - 2 \sin^2 \theta_W \\
  \beta_V^{(1)} &=& - 2 \sin^2 \theta_W .
\end{eqnarray}
\end{mathletters}
Since $\sin^2 \theta_W$ = 0.227, one has
$\beta_V^{(0)}\approx -\beta_V^{(1)}$; thus the isovector and the isoscalar
terms in (\ref{eq:RLAV}) almost cancel and $R^L_{AV}$ nearly vanishes
in an independent-particle model such as the relativistic Fermi
gas (RFG), as discussed in ref.~\cite{Don92}.
More specifically the $R^L_{AV}$ of the RFG is very small and negative
at $q=$ 300 MeV/c and then remains very small, but positive, up to quite
large momenta (say 2 GeV/c).

It is also clearly apparent that nuclear correlations
altering the isoscalar-isovector balance will markedly affect
$R^L_{AV}$. Such effects were first explored in the framework
of a pion-correlated RFG in our previous work \cite{Bar94}. There we
found a huge enhancement of $R^L_{AV}$
with respect to the pure RFG predictions for momentum transfers up to about
500 MeV/c. Moreover, we saw that in the pionic model $R^L_{AV}$ does not
keep the same sign as a function of $\omega$, but is strongly negative at
low frequencies, then becoming strongly positive at large frequencies.

Not surprisingly, these pion-induced features of the pv longitudinal
response bear important consequences for the asymmetry \cite{Don92}
\begin{equation}
  {\cal{A}}(\theta;q,\omega) =
    {\cal{A}}_0{v_L R^L_{AV}(q,\omega) + v_T R^T_{AV}(q,\omega) +
    v_T' R^{T'}_{VA}(q,\omega) \over v_L R^L(q,\omega) + v_T R^T(q,\omega)},
\label{eq:asym}
\end{equation}
which is measured in the inelastic scattering of longitudinally polarized
electrons. Indeed, when inserted into (\ref{eq:asym}), the $R^L_{AV}$ of
the pionic RFG leads, for the following observable (defined in
ref.~\cite{Bar94})
\begin{equation}
  \Delta{\cal{A}}(q,\theta) \equiv {1 \over{\Delta\omega}}\Bigl[
  \int_{\omega_{min}}^{\omega_{QEP}}\!\!d\omega
   {\cal{A}}(\theta;q,\omega) -
  \int_{\omega_{QEP}}^{\omega_{max}}\!\!d\omega
   {\cal{A}}(\theta;q,\omega)\Bigr]\ ,
\label{eq:delta_a}
\end{equation}
to values which are an order of magnitude larger than those obtained for the
pure RFG. This occurs for the same range of momentum transfers referred to
above and in forward-angle electron scattering.
In (\ref{eq:asym}) $R^T_{AV}$ and $R^{T'}_{VA}$ are the weak neutral
transverse and axial responses, respectively, whereas $R^T$ is the em
transverse response. Furthermore, $v_L, v_T$ and $v_T'$ are the usual
leptonic factors defined, for example, in ref.\cite{Don92}, while
\begin{equation}
  {\cal{A}}_0 = {G\left| Q^2\right|\over 2\pi\alpha\sqrt{2}}\simeq
    3.1\times 10^{-4}\tau
\label{eq:A0}
\end{equation}
sets the scale of the asymmetry; here $\tau=|Q^2|/(4m_N^2)$,
$\alpha$ is the electromagnetic and $G$ the Fermi coupling constant.
Finally the limits of the integrals over the asymmetry appearing in
(\ref{eq:delta_a}) are given by
\begin{mathletters}
\begin{eqnarray}
  \omega_{min} &=& \sqrt{(k_F-q)^2+m_N^2}-\sqrt{k_F^2+m_N^2} ,
    \label{eq:omegamin}\\
  \omega_{max} &=& \sqrt{(k_F+q)^2+m_N^2}-\sqrt{k_F^2+m_N^2}\
    \label{eq:omegamax}
\end{eqnarray}
\end{mathletters}
and
\begin{equation}
  \omega_{QEP}=|Q^2|/(2m_N) = 2 m_N\tau\ ,\quad
  \Delta\omega\equiv\omega_{max}-\omega_{min}\ .
\label{eq:peak}
\end{equation}

Prompted by the unusual sensitivity seen in these previous studies,
in the present paper we again
explore the parity-violating longitudinal response, always in the basic
framework of the RFG model, but with a much richer interaction than the
purely pionic one, namely the Bonn potential \cite{Mac87}. Indeed the latter
includes, besides the $\pi$, the $\rho, \sigma$ and $\omega$ mesons as well.
Furthermore, rather than performing the calculation in first-order
perturbation theory, as we did for the pionic case in
refs.~\cite{Alb93,Bar94}, we do it now in the framework of a Hartree-Fock
(HF) based fully antisymmetrized RPA. The motivation of the present work
is thus clear: we wish to test whether the features of $R^L_{AV}$
found previously survive in a broader many-body framework and when
a more realistic interaction is employed.

As in our previous work, here we also study the sensitivity of $R^L_{AV}$
obtained in our model to modified {\em nucleonic} dynamics by allowing for
the presence of strangeness in the electric form factor of the nucleon.

\section{Diagrammar}
\label{sec:diagrammar}

\begin{figure}[p]
\begin{center}
\mbox{\epsfig{file=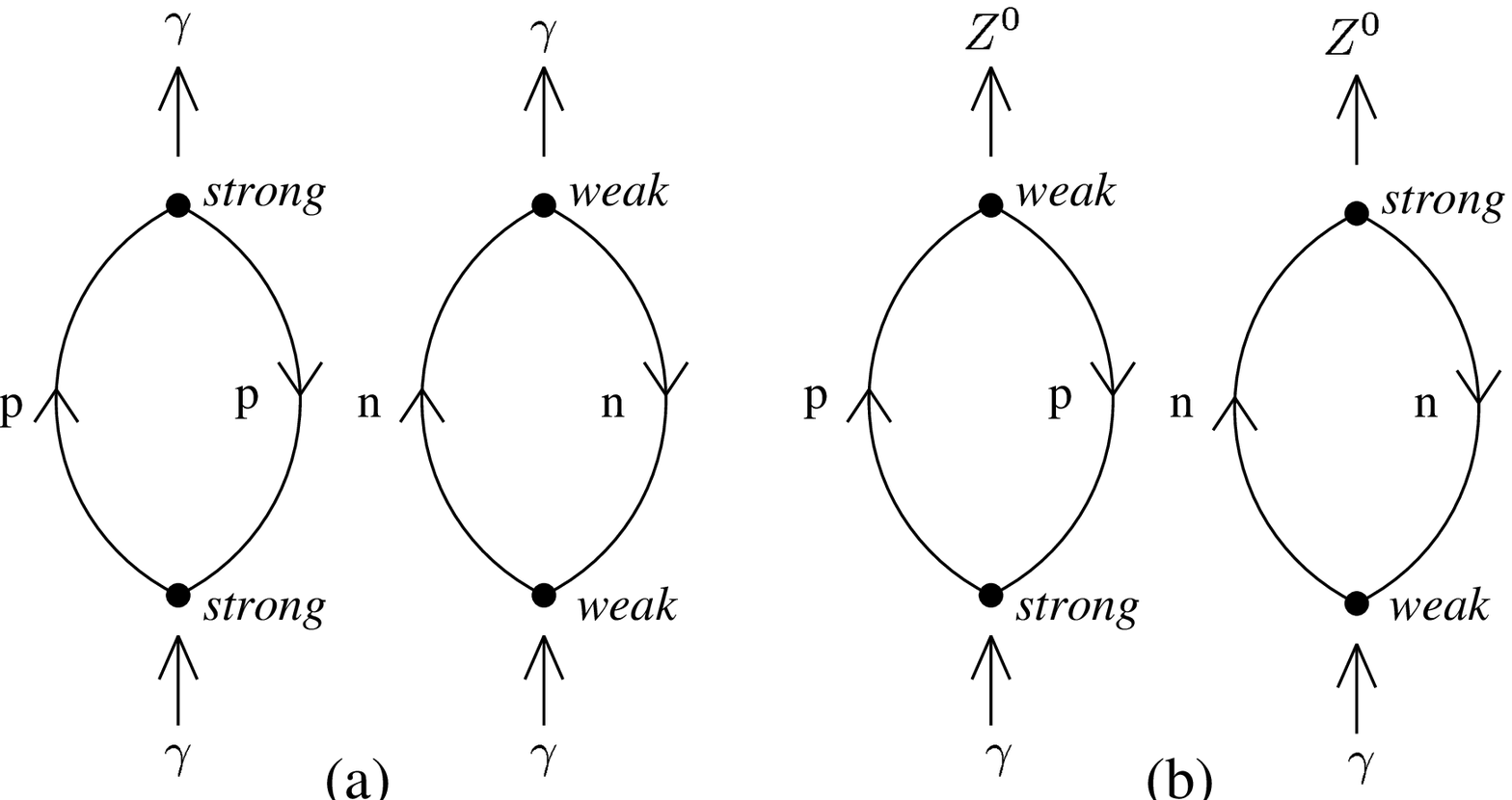,width=.59\textwidth}}
\vskip 2mm
\caption{
Feynman diagrams representing the free particle-hole polarization
propagator for the em (a) and pv (b) longitudinal response.
The excitation of proton (p) and neutron (n) particle-hole pairs is shown
separately. The labels {\em strong} and {\em weak} refer to the strength of
the nucleon coupling to the photon $\gamma$ or to the vector boson $Z^0$.
  }
\label{fig:diagrams-free}
\end{center}
\vfill
\begin{center}
\mbox{\epsfig{file=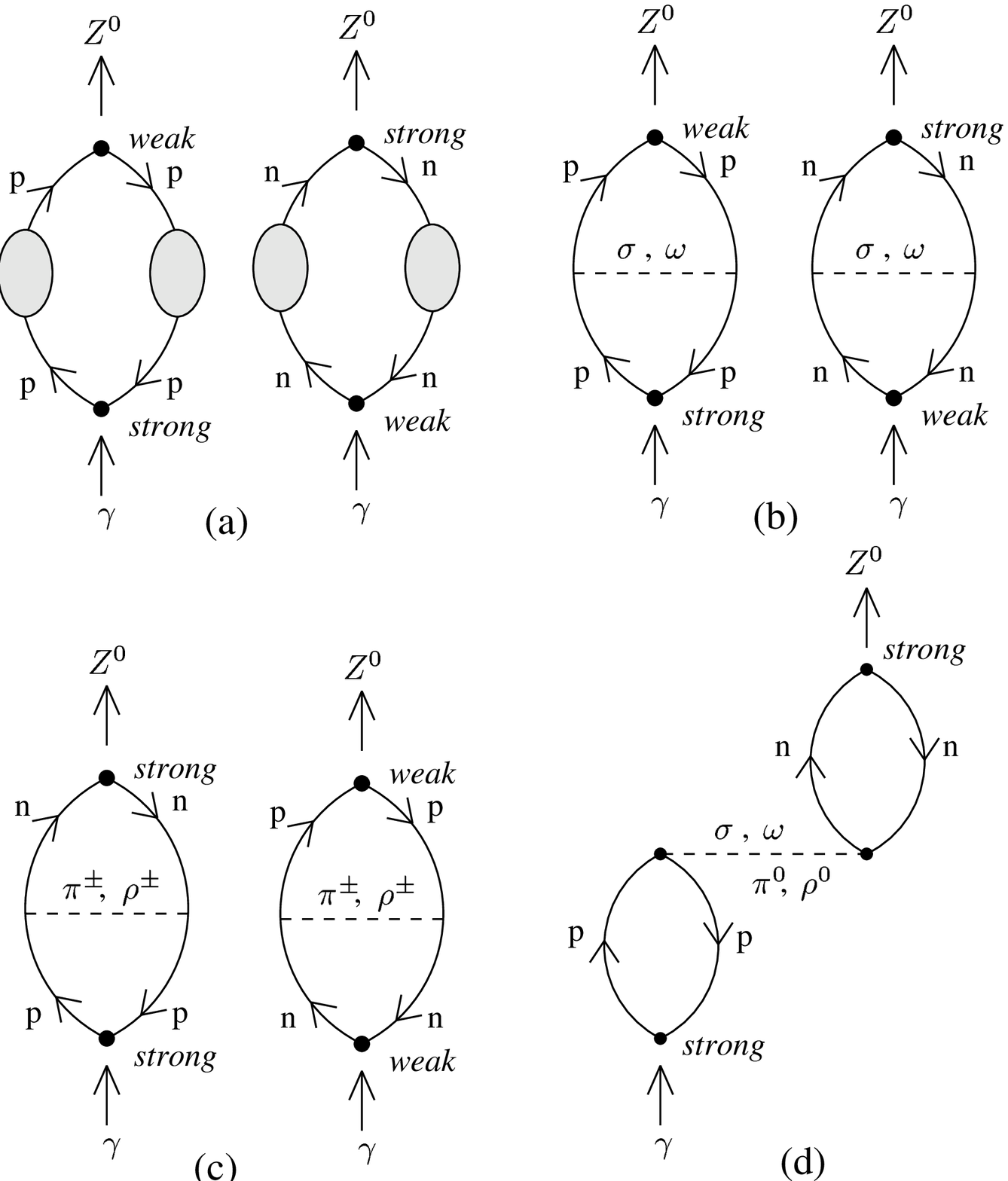,width=.59\textwidth}}
\vskip 2mm
\caption{
Feynman diagrams corresponding to the correlated pv longitudinal
response.
The meaning of the labels is as in Fig.~\protect{\ref{fig:diagrams-free}}.
The HF case is illustrated in (a), where the bubbles represent the nucleon
self-energy;
in (b) and (c) we show the first-order exchange diagrams induced by the
exchange of isoscalar and isovector mesons, respectively;
in (d) we illustrate one of the possible first-order ring diagrams involving
the exchange of neutral mesons.
  }
\label{fig:diagrams-corr}
\end{center}
\end{figure}

Before presenting detailed results using our model, we note that it is
possible to obtain an alternative, more physical picture of why $R^L_{AV}$
is small in the independent-particle model and why isospin-correlations
are so important in determining its ultimate size by expressing the
observables in a language which explicitly refers to neutrons and protons
rather than employing isospin labeling.
Indeed, by inspecting Fig.~\ref{fig:diagrams-free}, where the diagrams
describing both the em and the pv longitudinal responses for a free system
are displayed, one easily understands why already in the non-interacting
case the em longitudinal response turns out to be substantial: both of its
vertices can be large as they involve the coupling of a longitudinal photon
to a proton. In contrast, one of the vertices entering into the pv response
is {\em always} small, since either the longitudinal coupling of a photon to
a neutron or of a $Z^0$ to a proton is involved. This last occurrence is
often phrased by saying that the $Z^0$ is blind to protons. To quantify the
meaning of ``large'' and ``small'' we note that typically
the $\gamma -n$ coupling is about 1/10 that of the $\gamma -p$ and likewise,
but inversely, for the $Z^0 -p$ and $Z^0 -n$ couplings.

The above arguments clearly do not apply to the transverse response both
because in this case the $\tau=0$ channel is much weaker than the
$\tau=1$ channel, being essentially proportional to the squares of the very
different isoscalar and isovector magnetic moments of
the nucleon, and because a neutron can strongly couple to a photon via its
spin. The axial-vector response, also being essentially isovector,
similarly does not display the sensitivity expected for the longitudinal
response.

The question being addressed here is whether correlations among the nucleons
can change the above picture. In this connection, within the HF-RPA
framework, the following observations are in order:
\begin{itemize}
\item[i)]{the HF field clearly cannot enhance the pv response, since
{\em any} self-energy insertion along a nucleonic line never converts a
proton into a neutron or vice versa (see Fig.~\ref{fig:diagrams-corr}a);}
\item[ii)]{also the diagrams where {\em isoscalar} mesons are exchanged
between a particle and a hole which belong to the same ring cannot do it
for the same reason (Fig.~\ref{fig:diagrams-corr}b);}
\item[iii)]{however the diagram where charged {\em isovector} mesons are
exchanged between the particle and the hole of a ring
(Fig.~\ref{fig:diagrams-corr}c) {\em does} convert a neutron (proton) into a
proton (neutron) and it is precisely this diagram that was found to play the
crucial role in shaping the $R^L_{AV}$ of a pionic-correlated RFG and to
lead to a dramatic enhancement of the observable (\ref{eq:delta_a}), as
referred to in the Introduction;}
\item[iv)] {finally, the ring diagrams (Fig.~\ref{fig:diagrams-corr}d) can
always induce a neutron-proton (or vice versa) conversion (of course only
when the mesons carrying the force are not prevented from connecting the
rings by spin selection rules). Indeed each meson is either a scalar or a
vector in isospace and accordingly, in the ring scheme, will always alter
the balance between the $\tau=0$ and $\tau=1$ channels. For example, this
is the case for the $\sigma$ and $\omega$ mesons which, being isoscalar,
only act between $\tau=0$ rings.}
\end{itemize}

\section{Parity-Violating Longitudinal Response}
\label{sec:pv_long_resp}

\begin{figure}[p]
\begin{center}
\mbox{\epsfig{file=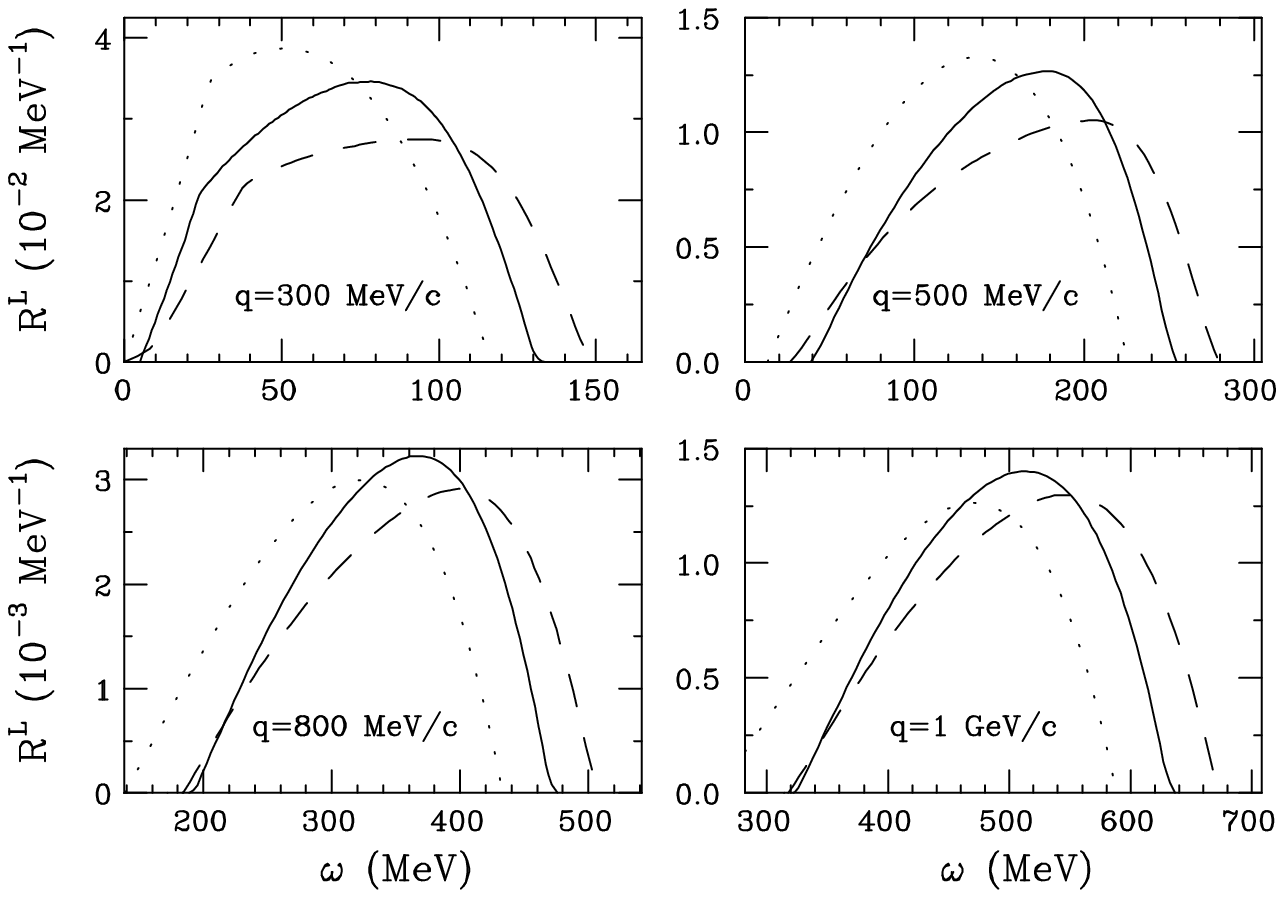,height=87mm}}
\caption{
The em longitudinal response $R^L$ is shown as a function of
$\omega$ at $q=$ 300, 500, 800 and 1000 MeV/c. The dotted curves correspond
to the free RFG calculation with $k_F=$ 225 MeV/c,
the dashed curves to the HF-RPA calculation
with $k_F=$ 225 MeV/c, the solid curves to HF-RPA with $k_F=$ 200 MeV/c.
  }
\label{fig:RL-HF-RPA}
\end{center}
\vfill
\begin{center}
\mbox{\epsfig{file=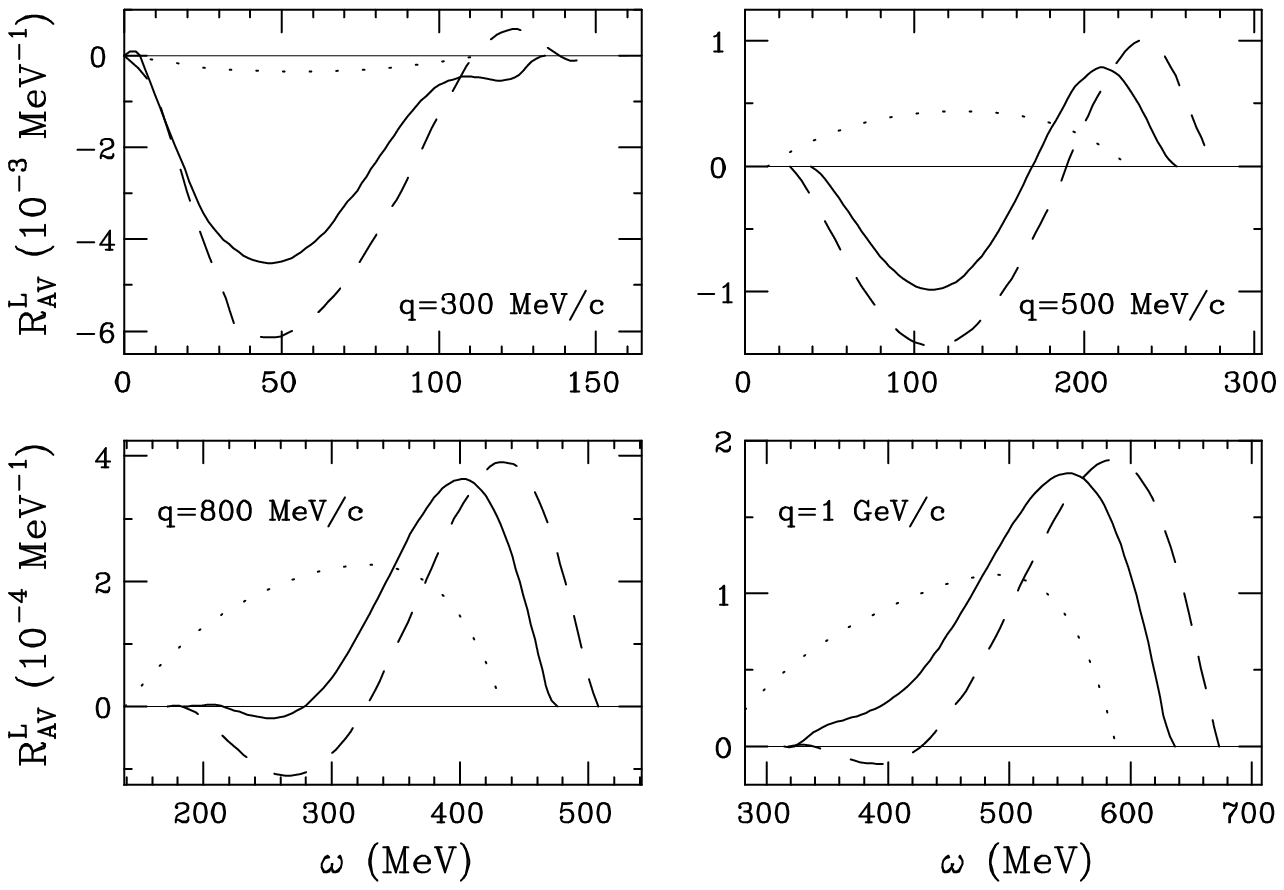,height=87mm}}
\caption{
The same as in Fig.~\protect{\ref{fig:RL-HF-RPA}} for the pv
longitudinal response $R^L_{AV}$.
  }
\label{fig:RLAV-HF-RPA}
\end{center}
\end{figure}

As already anticipated, in this section we discuss the pv longitudinal
response, $R^L_{AV}$, of the RFG in a HF based antisymmetrized RPA
framework, taking advantage of a calculation of the em charge response
we have recently performed \cite{Bar95a} utilizing as dynamical input the
simplest version of the Bonn potential \cite{Mac87}, which allows for the
exchange of the $\pi, \rho, \sigma$ and $\omega$ mesons between the nucleons.
For a thorough discussion of the relativistic and dynamical content of our
approach, see ref.~\cite{Bar95a}.
Concerning the Fermi momentum $k_F$ used, we adopt the procedure of
ref.~\cite{Bar95a} where it was found that one should
associate the value $k_F\approx 200$ MeV/c
to $^{12}$C (actually $k_F=225$ MeV/c is required to reproduce the
experimental width of the inclusive cross section in $^{12}$C {\em in a pure
Fermi gas framework}, whereas such a value would correspond to a
medium-heavy nucleus in our correlated model \cite{Bar95a}).

In Figs.~\ref{fig:RL-HF-RPA} and \ref{fig:RLAV-HF-RPA} we display the em
and pv longitudinal responses in the full HF-RPA model. The charge em
response has already been discussed at length in \cite{Bar95a}:
suffice to note again here the hardening that occurs up to the largest
considered momenta (namely 1 GeV/c) and the quenching that is seen up to
about 500 MeV/c with respect to the free case. The former stems essentially
from the HF field, mainly through the roles played by the $\sigma$ and
$\omega$ mesons, whereas the crucial role of the pion in inducing the
latter was emphasized in ref.~\cite{Bar95a}. Of significance is also the
growing importance of these effects in going from light to medium-heavy
nuclei.

Turning to the weak neutral longitudinal response, its large enhancement
with respect to the free case is fully apparent from the results
presented in Fig.~\ref{fig:RLAV-HF-RPA}. Thus the basic sensitivity of this
observable to effects that go beyond the mean field is confirmed. At
around 500 MeV/c the oscillatory behaviour in frequency of the weak neutral
response seen in the figure becomes pronounced. This effect is then seen to
fade away at larger momenta, where $R^L_{AV}$, although still hardened and
somewhat enhanced, is no longer so strikingly different from the RFG case.

\begin{figure}[p]
\begin{center}
\mbox{\epsfig{file=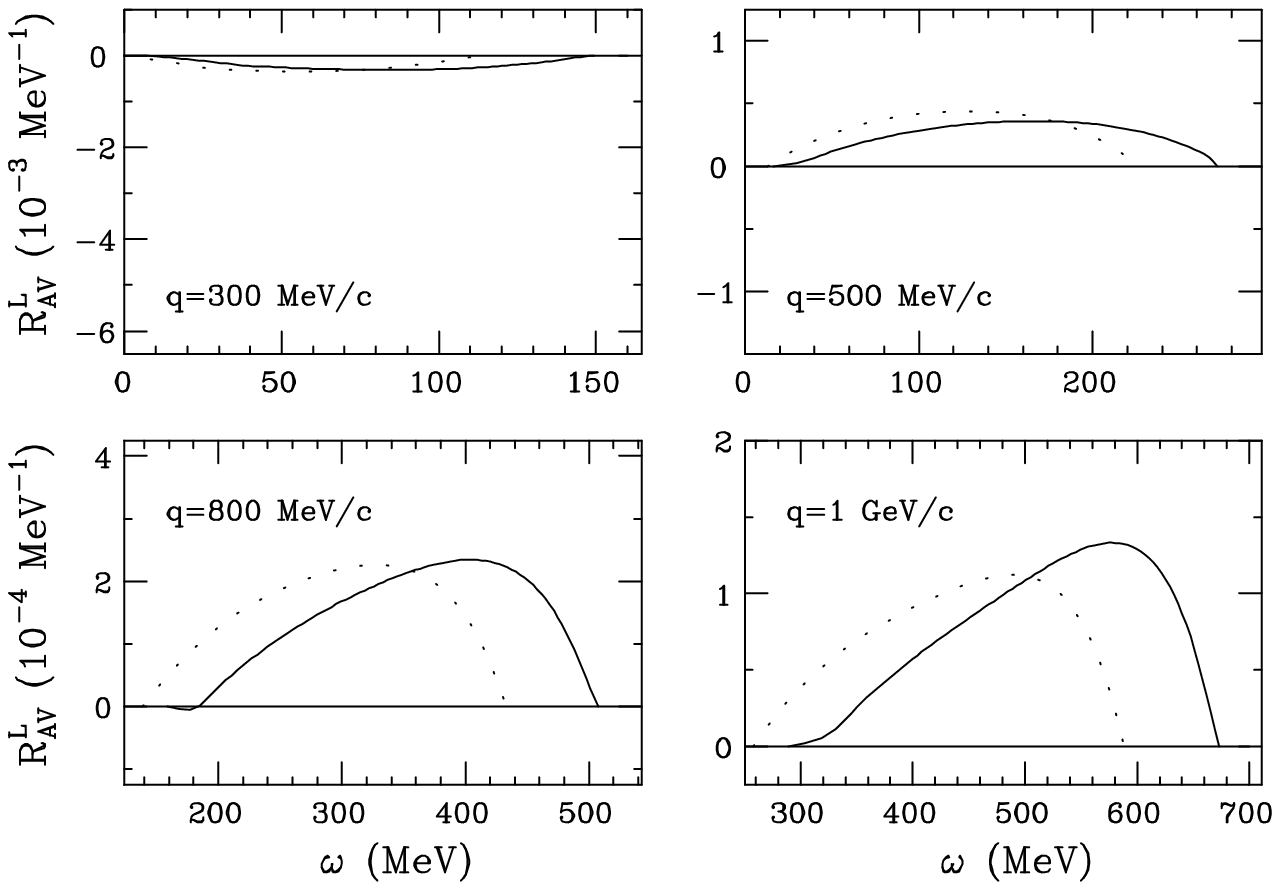,height=87mm}}
\caption{
The free (dotted) and HF (solid) pv longitudinal response $R^L_{AV}$
versus $\omega$ at $q=$ 300, 500, 800 and 1000 MeV/c;
$k_F=$ 225 MeV/c for the free RFG and $k_F=$ 200 MeV/c for the correlated
response.
  }
\label{fig:RLAV-HF}
\end{center}
\vfill
\begin{center}
\mbox{\epsfig{file=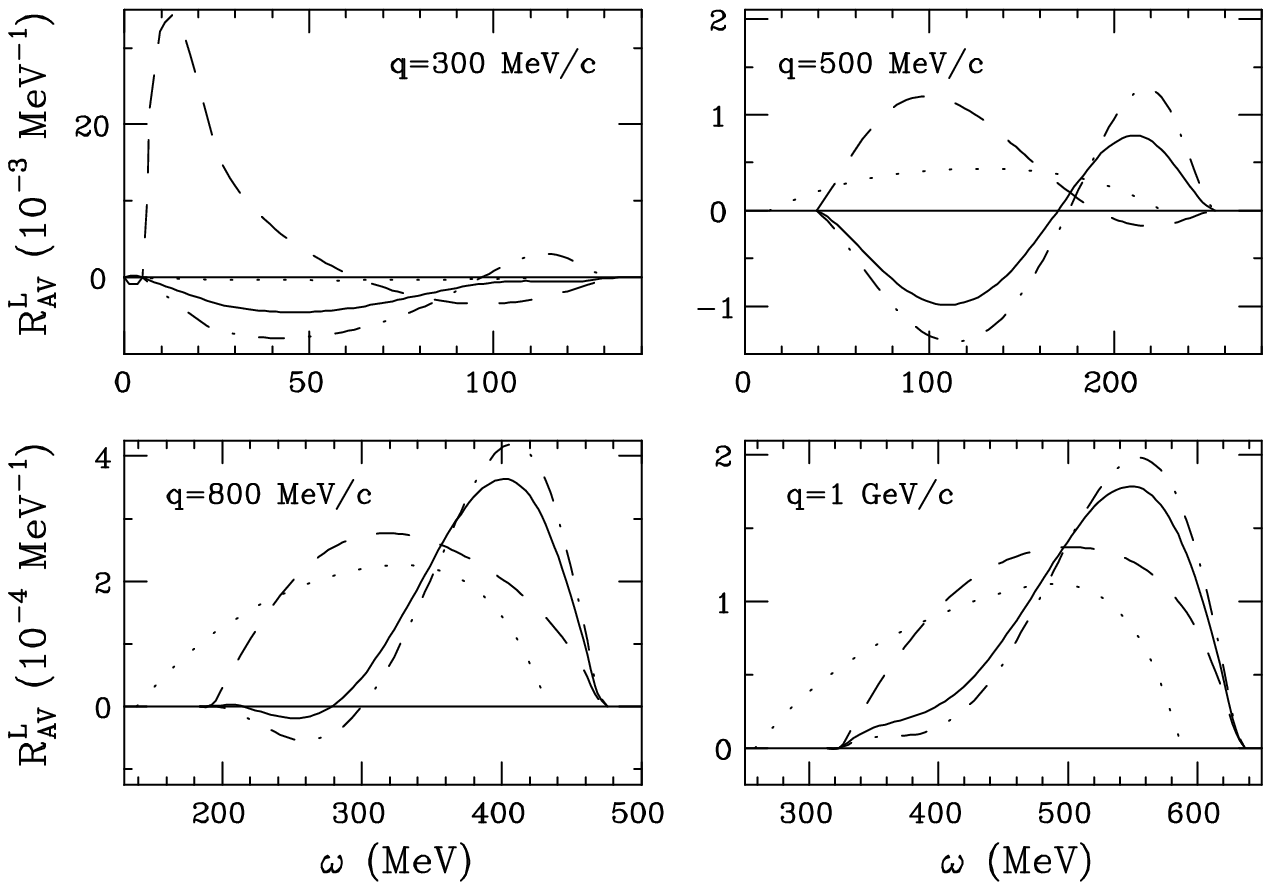,height=87mm}}
\caption{
The pv longitudinal response versus $\omega$ at $q=$ 300, 500, 800 and
1000 MeV/c. The dotted lines correspond to the free RFG case, the solid
lines to the HF-RPA calculation; the dashed lines represent the pure ring
approximation, whereas the dot-dashed ones the pure exchange contribution.
$k_F=$ 225 MeV/c for the free RFG and $k_F=$ 200 MeV/c for the correlated
response.
  }
\label{fig:RLAV-HF-ring-exch-rpa}
\end{center}
\end{figure}

In order to understand the dynamical origin of these results, let us examine
the various ingredients that go into our model, starting with
the HF field alone. As mentioned above, the mean field leads to a
strong hardening of the em antisymmetrized HF-RPA charge response with
respect to the pure RFG response for transferred momenta up to about
1 GeV/c. In spite of this and in line with the considerations of the
previous section, we see in Fig.~\ref{fig:RLAV-HF} that $R^{L,HF}_{AV}$
is almost identical to the $R^L_{AV}$ of the RFG, but for the shift in the
response region (induced by the HF field itself \cite{Bar95a}).
This result shows how selective $R^L_{AV}$ can be
with respect to the many-body scheme employed for its evaluation: in
particular, it filters out almost completely the mean field. Also noteworthy
is that the sign inversion of the pv longitudinal response \cite{Don92},
occurring for a momentum transfers between 300 and 500 MeV/c, remains
essentially unmodified in going from the free RFG case to the HF one.

Turning to a non-symmetrized RPA scheme built on a HF basis (namely to
consider ring diagrams only) we observe in Fig.~\ref{fig:RLAV-HF-ring-exch-rpa}
the huge impact the $\sigma$ and $\omega$ mesons have on the pv
longitudinal response via the ring diagrams. This persists up to about
500 MeV/c and reflects the dominance of the attractive $\sigma$ over
the repulsive $\omega$. It then fades away except for the usual shift
induced by the HF field. Insofar as the ring diagrams are concerned,
the $\sigma$ and $\omega$ affect  the pv response more than the em charge
response, a signature of the enhanced sensitivity of the former to selected
classes of diagrams.

\begin{figure}[p]
\begin{center}
\mbox{\epsfig{file=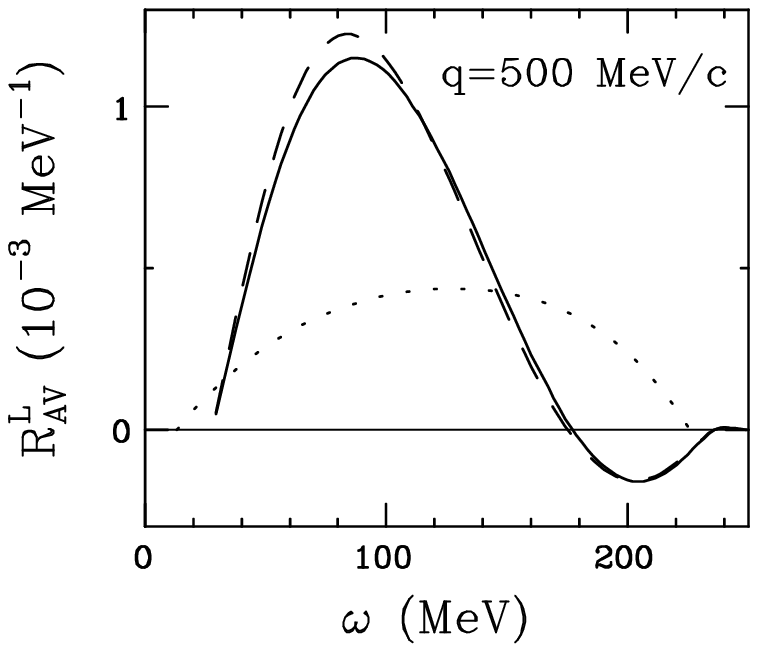}}
\vskip 2mm
\caption{
The pv longitudinal response is displayed as a function of $\omega$
at $q=$ 500 MeV/c. The dotted curve corresponds to the free RFG case, the
solid (dashed) curve to the HF-RPA (ring) approximation when only the
$\sigma$ and $\omega$ mesons are included in the nucleon-nucleon interaction;
$k_F=$ 225 MeV/c for the free RFG and $k_F=$ 200 MeV/c for the correlated
response.
  }
\label{fig:RLAV-sigma-omega}
\end{center}
\vfill
\begin{center}
\mbox{\epsfig{file=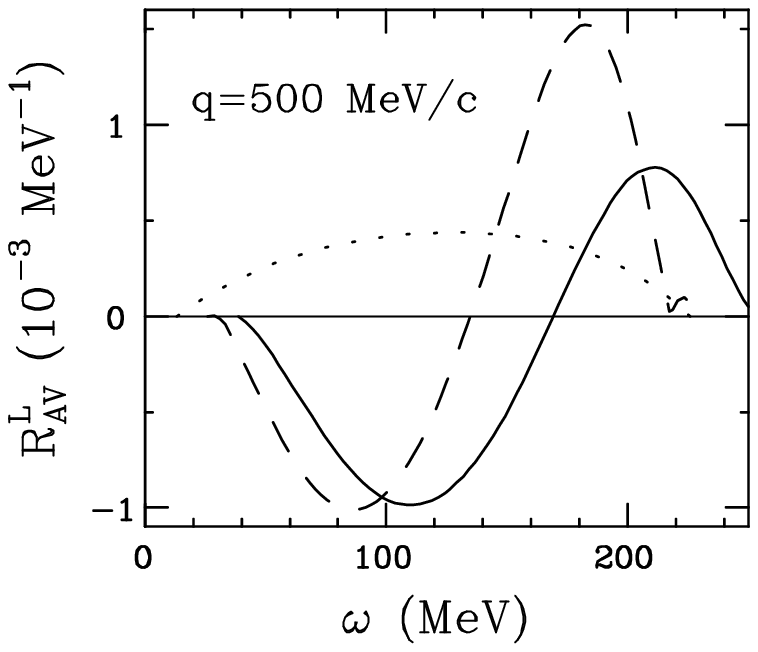}}
\vskip 2mm
\caption{
The pv longitudinal response is shown as a function of $\omega$ at
$q=$ 500 MeV/c. The dotted curve corresponds to the free RFG case, the solid
curve to the HF-RPA approximation with the full Bonn potential, the dashed
curve
is obtained in the HF-RPA approximation with a pure pion-exchange interaction;
$k_F=$ 225 MeV/c for the free RFG and $k_F=$ 200 MeV/c for the correlated
response.
  }
\label{fig:RLAV-pion}
\end{center}
\end{figure}

As argued in sect.~\ref{sec:diagrammar} and already seen
to occur in relation to the self-energy insertions, the pv longitudinal
response can be most effective in filtering out certain diagrams.
In this connection another example is offered by the exchange diagrams
(Fig.~\ref{fig:diagrams-corr}b) involving the {\em isoscalar} $\sigma$ and
$\omega$ mesons. In fact, the switching on of these effects leaves
$R^L_{AV}$ almost unaltered with respect to the results obtained in the
HF-ring scheme, as illustrated in Fig.~\ref{fig:RLAV-sigma-omega} for the
typical value $q=$ 500 MeV/c. In contrast, when considering $R^L$, it was
shown in ref.\cite{Bar95a} that the exchange term associated with the
$\sigma$ and $\omega$ is quite important in counteracting the contribution
of the same mesons arising from the ring diagrams (actually, in
ref.~\cite{Bar95a} it was seen that the
exchange diagrams dominate over the ring contributions).

Of the various effects embodied in $R^L_{AV}$ those brought about by the
pion are special and require further discussion. The pion contributes to
both the em and the pv longitudinal responses essentially only via
the exchange term (leaving aside a small Fock contribution to the
self-energy). However, when the pion is allowed to enter the exchange
diagrams {\em only simultaneously} with the other mesons but not alone,
or in other words, only by {\em interfering} with the $\sigma$ and
$\omega$ (and, to a lesser extent, with the $\rho$), then it cancels out
almost completely the previously discussed ring effect stemming from the
$\sigma$ and $\omega$. This outcome is clearly apparent in
Fig.~\ref{fig:RLAV-HF-ring-exch-rpa} (note that the $\rho$ contribution is
quite
small) and goes in parallel with the findings of
ref.~\cite{Bar95a}, where the most effective action of the pion on the em
charge response was found to arise precisely from its interference with the
other mesons.

Analogously, in the pv case the contribution arising from the interference
of the $\pi$ with the other mesons in the exchange diagrams is large indeed,
in that it serves to wash out almost completely the ring contribution.
As a consequence what remains is essentially the contribution of the
pion alone to $R^L_{AV}$.
At variance with the em case where the relevance of the pure pionic
contribution was found to be minor with respect to that of the other
mesons, in the pv case, owing both to the previously discussed cancelling
and filtering of contributions and to the imbalance induced by the pion
between the two isospin channels, we see a significant alteration of the
shape of $R^L_{AV}$ with respect to the free case.
Actually, to illustrate how similar the pv longitudinal responses calculated
with the pion only and with the Bonn potential are, we display both of
them in Fig.~\ref{fig:RLAV-pion} for q=500 MeV/c.
The results for the Bonn potential over the whole range of momenta
are instead illustrated in Fig.~\ref{fig:RLAV-HF-RPA}, where the marked
dependence upon $k_F$ is also shown.

\begin{figure}[tb]
\begin{center}
\mbox{\epsfig{file=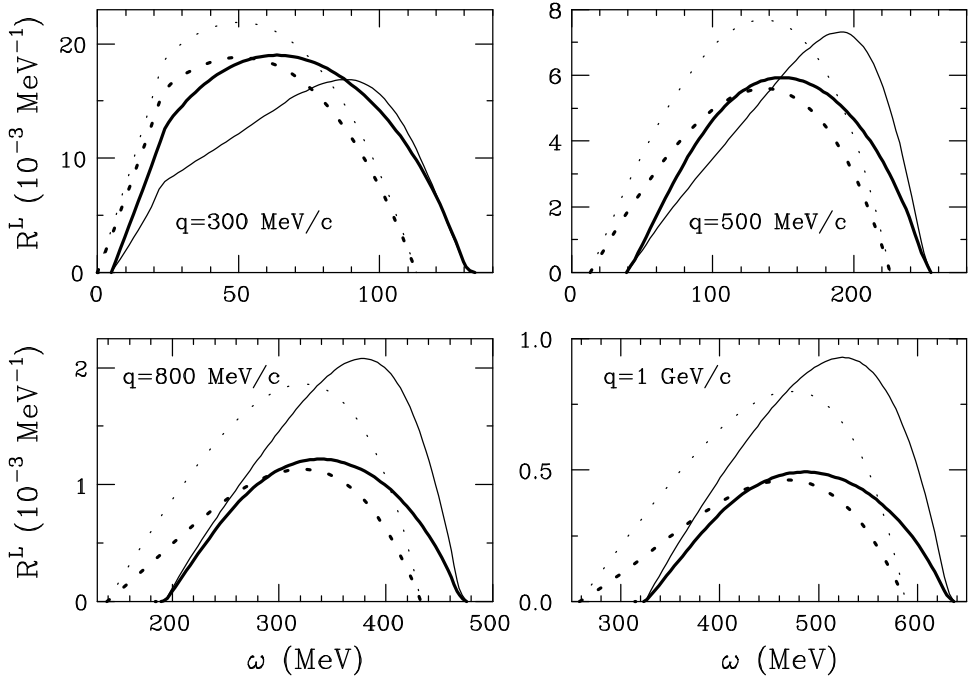}}
\caption{
The isoscalar (thin curves) and isovector (thick curves) em
longitudinal responses, as defined in eq.~(\protect{\ref{eq:RL01}}), are
displayed as functions of $\omega$ at $q=$ 300, 500, 800 and 1000 MeV/c.
The dotted curves are the free RFG results, the solid curves correspond to
the HF-RPA calculation;
$k_F=$ 225 MeV/c for the free RFG and $k_F=$ 200 MeV/c for the correlated
response.
  }
\label{fig:RL-isospin}
\end{center}
\end{figure}

Worth noticing is the following (somewhat unexpected) result:
the $R^L_{AV}$ obtained within the antisymmetrized HF-RPA scheme with a Bonn
potential turns out in the end not to be much different from the one
obtained in first-order perturbation theory with the pion alone. Indeed the
crucial exchange diagram for the pion was shown in ref.~\cite{Bar94} to be
quite accurately treated
already in first-order.

Finally, to get a direct assessment of the isoscalar-isovector imbalance,
which lies at the core of Fig.~\ref{fig:RLAV-HF-RPA}, we display in
Fig.~\ref{fig:RL-isospin} the separated $\tau=1$ and $\tau=0$ longitudinal
responses:
\begin{mathletters}
\label{eq:RL01}
\begin{eqnarray}
  R^L(\tau=0) &=& { \beta_V^{(1)} R^L - R^L_{AV} \over
  \beta_V^{(1)}-\beta_V^{(0)}}\\
  R^L(\tau=1) &=& { \beta_V^{(0)} R^L - R^L_{AV} \over
  \beta_V^{(0)}-\beta_V^{(1)}}.
\end{eqnarray}
\end{mathletters}
In the figure one fully appreciates the extent of the disruption induced
by the
nuclear correlations on the isoscalar RFG response (particularly at not too
high momentum transfers), while their influence in the isovector channel
appears to be mild.

\section{Strangeness}
\label{sec:strangeness}

In addition to the intrinsic interest we have in $R^L_{AV}$ as an observable
that reflects the many-body content of the nuclear problem in a novel way,
the pv longitudinal response may also provide important clues on nucleonic
structure itself. Specifically, as in our previous work
\cite{Don92,Bar94,Mus94}, we
consider how the results presented here would be affected by the presence
of a nonzero electric strangeness form factor, $G_E^{(s)}$, parametrized as
\begin{equation}
  G_E^{(s)}(\tau) = \rho_s {\tau\over \left[1+\lambda_D^V\tau\right]^2}
    {1\over\left[1+\lambda_E^{(s)}\tau\right]} ,
\label{eq:Galster}
\end{equation}
with $\lambda_D^V=4.97$. For the constants entering into (\ref{eq:Galster}),
in past work the following three choices have been considered:
\begin{mathletters}
\label{eq:strange}
\begin{eqnarray}
  \rho_s &=& 0
\label{eq:strange1} \\
  \rho_s &=& -3\, ,\quad\lambda_E^{(s)}=5.6
\label{eq:strange2} \\
  \rho_s &=& -3\, , \quad\lambda_E^{(s)}=0 \,.
\label{eq:strange3}
\end{eqnarray}
\end{mathletters}
The first one corresponds to the absence of strangeness, the second one to
the case where electric strangeness behaves much like $G_{En}$, whereas
the third one describes the situation in which the electric strangeness
behaviour is very significant at high-$|Q^2|$, being uninhibited by the
Galster-like factor containing $\lambda_E^{(s)}$.

\begin{figure}[tb]
\begin{center}
\mbox{\epsfig{file=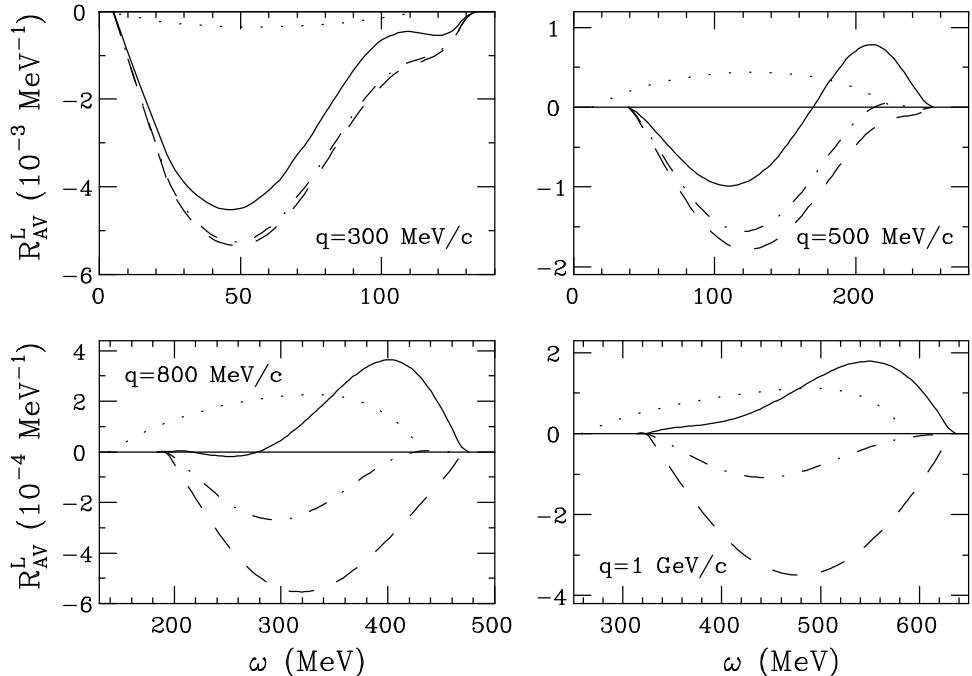}}
\caption{
The pv longitudinal response is displayed versus $\omega$ at
$q=$ 300, 500, 800 and 1000 MeV/c.
The dotted curve refers to the free RFG case, the solid, dot-dashed and
dashed curves represent the HF-RPA results with $G_E^{(s)}$ given by the
three models of eq.~(\protect{\ref{eq:strange}}), respectively.
$k_F=$ 225 MeV/c for the free RFG and $k_F=$ 200 MeV/c for the correlated
response.
  }
\label{fig:RLAV-strangeness}
\end{center}
\end{figure}

In Fig.~\ref{fig:RLAV-strangeness} we present our results for $R^L_{AV}$,
again obtained for the RFG in the antisymmetrized HF-RPA scheme with the
Bonn potential, now with the addition of $G_E^{(s)}$. We see clearly
emerging from this the fact that
the pv longitudinal response with $G_E^{(s)}$ included is significantly
different from the $R^L_{AV}$ computed with no strangeness at all once
the momentum transfer is large enough to allow this form factor, which
is suppressed at low-$|Q^2|$ by the required factor of $\tau$ in the
numerator of (\ref{eq:Galster}), to enter at all.

\section{Concluding Remarks}
\label{sec:conclusions}

In the present work we have revisited the issue of using the pv
longitudinal response as a tool to study, on the one hand, the many-body
dynamics underlying responses in general in the quasielastic region, and,
on the other, the role played by electric strangeness in the nucleon in
producing this response.  Within the context of a model that incorporates
HF-RPA correlations in a relativistic Fermi gas basis we have explored the
roles played by the $\pi, \rho, \sigma$ and $\omega$ mesons as they enter
via a version of the Bonn potential.

One goal of this study was to determine the extent to which the special
sensitivity of $R_{AV}^L$ seen in our previous work to particular classes
of isospin-correlations would survive in the presence of a strong
$\sigma$--$\omega$ mediated mean field. As discussed in detail in this
paper, we clearly confirm the previous expectation that the longitudinal
pv response, even in such a more sophisticated strong-coupled model as
that considered here, can be used to meter special classes of many-body
effects and, perhaps, at high enough momentum transfer to provide an
alternative way to study $G_E^{(s)}$.

\end{document}